\PassOptionsToPackage{numbers,sort&compress}{natbib}
\documentclass[preprint,12pt]{elsarticle}
\usepackage[total={6in, 8in}]{geometry}
\usepackage{amssymb}
\usepackage{caption} 

\makeatletter
\def\ps@pprintTitle{%
   \let\@oddhead\@empty
   \let\@evenhead\@empty
   \def\@oddfoot{\hfill\thepage\hfill}
   \let\@evenfoot\@oddfoot}
\makeatother

\begin{document}

\begin{frontmatter}

\title{Surface Orientation-dependent Corrosion Behavior of NiCr Alloys in Molten FLiNaK Salt}

\author[1]{Hamdy Arkoub}
\author[1]{Daniel Flynn}
\author[2]{Adri C.T. van Duin}
\author[1]{Miaomiao Jin\corref{cor}}
\affiliation[1]{organization={Department of Nuclear Engineering, The Pennsylvania State University},
            city={University Park},
            postcode={16802}, 
            state={PA},
            country={USA}}
\affiliation[2]{organization={Department of Mechanical Engineering, The Pennsylvania State University},
            city={University Park},
            postcode={16802}, 
            state={PA},
            country={USA}}
\cortext[cor]{Corresponding author: mmjin@psu.edu}

\begin{abstract}
%% Text of abstract

The corrosion behavior of NiCr alloys in molten FLiNaK salt is governed by complex Cr-F chemical interactions, necessitating a fundamental understanding for enhancing alloy performance in harsh environments. However, significant gaps remain in our understanding of the dynamic atomic-scale processes driving the progression of molten salt corrosion. This study employs reactive force field-based molecular dynamics simulations to unravel the influence of crystallographic orientation, temperature, and external electric fields on corrosion kinetics. The (100), (110), and (111) orientations of Ni$\mathrm{_{0.75}}$Cr$\mathrm{_{0.25}}$ alloys are evaluated at temperatures from 600 to 800°C, with and without electric fields. Results reveal that Cr dissolution and near-surface diffusion drive pitting-like surface morphology evolution. The (110) surface shows the highest corrosion susceptibility, while the (100) and (111) surfaces exhibit greater resistance, with (111) being the most stable. The corrosion activation energy, derived from the Arrhenius relation, ranges from 0.27 eV to 0.41 eV, aligning well with limited experimental data yet significantly lower than bulk diffusion barriers. This finding indicates that corrosion progression is primarily a kinetically controlled near-surface process, rather than being limited by bulk diffusion as suggested in previous understanding. Additionally, electric fields perpendicular to the interface are found to asymmetrically modulate corrosion dynamics, where a positive field (+0.10 V/Å) promotes Cr dissolution. In comparison, a negative field (-0.10 V/Å) largely suppresses corrosion, which can be effectively used to mitigate corrosion. These findings, along with atomistic details into the corrosion mechanisms, offer strategic perspectives for designing corrosion-resistant materials in advanced high-temperature molten salt applications.
\end{abstract}


\end{frontmatter}

%% main text
\section{Introduction}
\label{sec:sample1}

Molten salt corrosion poses a critical challenge in high-temperature applications across various industries, including energy systems, aerospace, and chemical processing \cite{roper2022molten,leblanc2010molten,kelly2014generation}. Molten fluoride salts, such as FLiNaK, are widely used due to their excellent thermal properties and chemical stability  \cite{williams2006assessment,delpech2010molten}. However, the highly corrosive nature of these salts at elevated temperatures leads to material degradation, complicating the selection of structural materials for long-term operation \cite{guo2018corrosion}. Unlike aqueous environments where passivating oxide layers can mitigate corrosion, molten salts destabilize these layers, exposing bare metals to direct attack \cite{sohal2010engineering}. Nickel-based alloys are among the most promising materials for molten salt environments due to their favorable high-temperature stability, mechanical strength, and corrosion resistance \cite{yvon2009structural,rosenthal1972development}. However, the strong affinity of fluorine for alloying elements, particularly Cr, promotes its segregation and dissolution, forming species such as $\mathrm{CrF_2}$ and $\mathrm{CrF_3}$ in the salt \cite{olson2009materials,ren2016adsorption,yin2018first,chan2022insights}. Such selective dissolution of constituent elements remains a major concern, as it alters surface chemistry and microstructural stability \cite{sohal2010engineering,williams2006assessment,olson2009materials,ouyang2013effect,leong2023kinetics}. 

Corrosion is governed by a complex coupling of interfacial reactions and diffusion processes. A growing body of research indicates that metal crystallographic orientation is critical in affecting corrosion susceptibility. For example, Chen et al. \cite{chen2024effect} demonstrated that the (100)-oriented aluminum exhibited superior corrosion resistance compared to (110) and (111) surfaces in a NaCl solution. Similarly, Wei et al. \cite{wei2022comparing} found that the corrosion resistance of Ni-based superalloys exposed to molten Na\(_2\)SO\(_4\)-NaCl salt at 750 °C followed the trend (110) $<$ (100) $<$ (210). These experimental findings highlight the need to understand the atomic-scale interactions across different crystallographic orientations, which can be effectively addressed through theoretical and computational approaches. First-principles methods, particularly density functional theory (DFT), have been used to study the early stages of Ni-based alloy corrosion in molten fluoride salts \cite{ren2016adsorption,yin2018first,startt2021ab,arkoub2024first}. However, their computational cost limits their applicability to small system sizes and short timescales, making them insufficient for capturing dynamic corrosion processes. Alternatively, reactive molecular dynamics (RMD) simulations using reactive force fields, such as ReaxFF, offer a more scalable approach \cite{van2001reaxff,senftle2016reaxff}. ReaxFF enables large-scale simulations of reactive events at solid-liquid interfaces, allowing for the investigation of chemical reactions and transport phenomena critical to understanding corrosion mechanisms \cite{senftle2016reaxff}. Several studies have demonstrated the effectiveness of ReaxFF-based MD simulations in exploring orientation-dependent corrosion behavior. For example, Subbaraman et al. \cite{subbaraman2013atomistic} investigated Fe oxidation and found that oxide growth was most pronounced on Fe(110), followed by Fe(111), with Fe(100) exhibiting the lowest oxidation rate. Huang et al. \cite{huang2022atomic} showed that surface orientation significantly affects oxidation rates in supercritical water, with Fe(110) corroding faster than Fe(100). Jeon et al. \cite{jeon2013nanoscale} further confirmed that Fe(110) surfaces had the highest oxidation rates. This increased susceptibility of the (110) surface to corrosion was attributed to its higher density of reactive sites \cite{jeon2013nanoscale} and the lower activation energy barrier for oxidation \cite{subbaraman2013atomistic}. 

Beyond intrinsic crystallographic effects, external electric fields serve as another critical factor in modulating corrosion behavior, which has also been explored using RMD simulations. Assowe et al. \cite{assowe2012reactive,assowe2012reactive2} investigated nickel oxidation under applied electric fields across (100), (110), and (111) orientations; it was shown that the electric field triggers nickel oxidation as water molecules form a bilayer structure on the metal surface.  Jeon et al. \cite{jeon2013nanoscale} found that electric fields accelerated the oxidation of Fe by enhancing oxygen migration through interstitial sites. Additionally, Zhou et al. \cite{zhou2022corrosion} showed that electric field can increase the oxidation rates of Fe(100) in supercritical CO$_2$ by 1.21 times compared to a field-free case, whereas reversing the field direction suppressed corrosion by stabilizing surface Fe atoms. In the molten salt corrosion system, how crystallographic orientation influences these reactions under external fields remains largely unexplored.  

\sloppy
The ability of RMD simulations to capture the interplay between crystallographic orientation and electric fields provides a powerful tool for understanding high-temperature corrosion mechanisms. Recent ReaxFF-based RMD simulations of the NiCr-FLiNaK system \cite{arkoub2024reactive} confirmed that Cr dissolution is primarily driven by strong Cr–F interactions, with increased fluoride coverage accelerating metal loss. However, the temperature dependence and the combined effects of crystallographic orientation and electric fields on molten salt corrosion under high-temperature conditions remain to be elucidated. To address this gap, in this study, we employ RMD simulations based on ReaxFF \cite{arkoub2024reactive} to systematically investigate how crystallographic orientation, temperature, and electric fields influence the corrosion behavior of Ni$\mathrm{_{0.75}}$Cr$\mathrm{_{0.25}}$ alloys in FLiNaK salt. By examining (100), (110), and (111) surface orientations across temperatures ranging from 600–800 °C under varying electric field conditions, we find that the (110) orientation exhibits the highest corrosion susceptibility compared to (100) and (111). Temperature-dependent simulations suggest that corrosion is primarily governed by the near-surface dissolution and diffusion of constituent elements rather than the bulk diffusion of Cr. Moreover, the application of an electric field proves to be an effective strategy for controlling the corrosion rate but has complications in modifying corrosion dynamics.  

\section{Methods}

RMD simulations were performed using the Amsterdam Modeling Suite (AMS) \cite{van2001reaxff, chenoweth2008reaxff, reaxff2022.1scm} with the ReaxFF force field developed for FLiNaK-NiCr system \cite{van2001reaxff, senftle2016reaxff,arkoub2024reactive}. Such force field was demonstrated to accurately reproduce the salt properties and corrosion behavior of NiCr alloys in FLiNaK molten salts \cite{arkoub2024reactive}. ReaxFF, based on a bond-order formalism, bridges the gap between first-principles approaches like DFT and classical molecular dynamics, enabling the modeling of reactive systems. A detailed description of ReaxFF can be found in \cite{van2001reaxff, chenoweth2008reaxff, senftle2016reaxff}.  

\begin{figure}[!ht]
	\centering
	\includegraphics[width=1.0\textwidth]{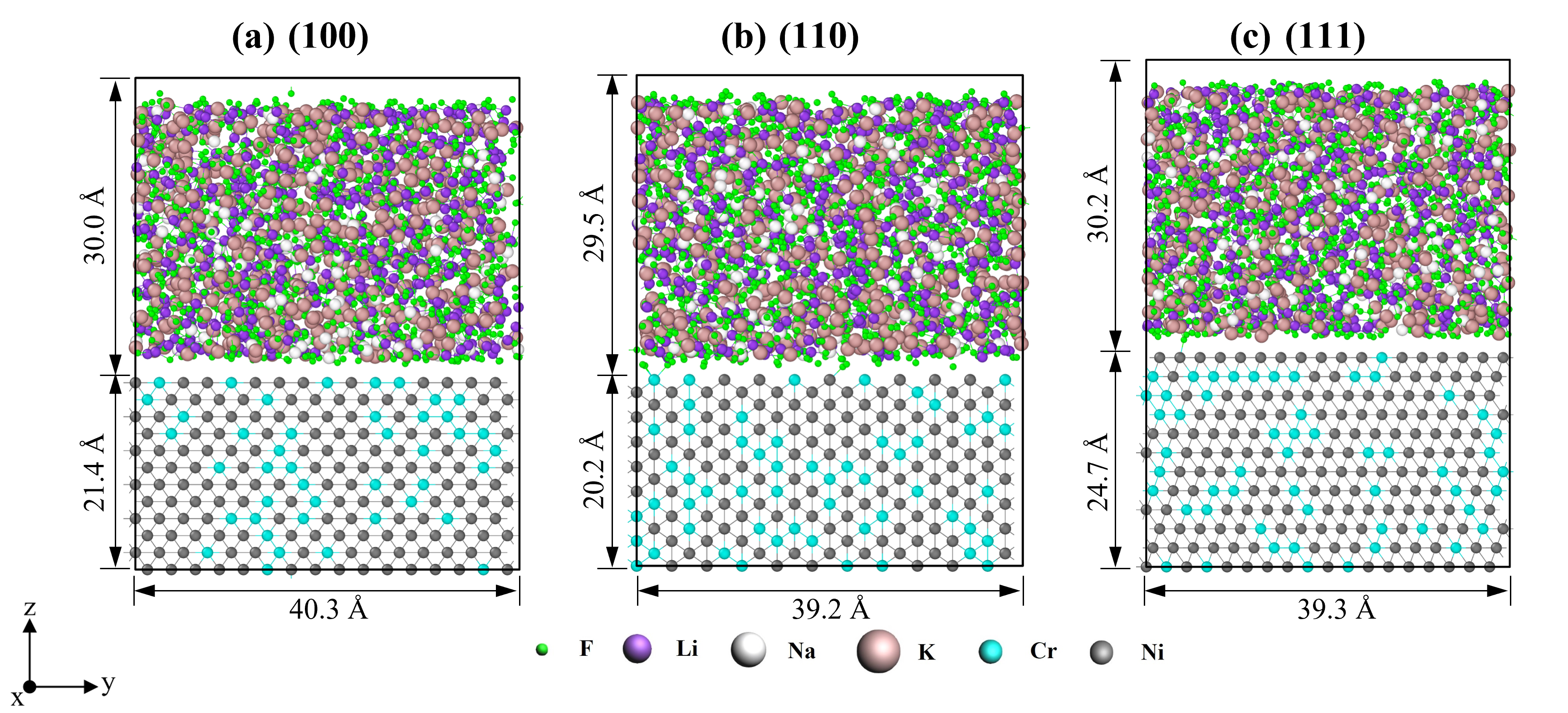}
	\caption{Side view of the initial simulation cells at 700 $\mathrm{^o}$C: Ni$\mathrm{_{0.75}}$Cr$\mathrm{_{0.25}}$ (a) (100), (b) (110), and (c) (111) slabs and molten FLiNaK salt at its equilibrium density for this temperature. The $x$ dimension is around 40.3 \AA.}
	\label{fig:initial}
\end{figure}

To investigate the effect of crystallographic orientation on corrosion behavior, we constructed FCC Ni\(_{0.75}\)Cr\(_{0.25}\) slabs with low-index (100), (110), and (111) surface orientations, respectively. The alloy contains 25\% Cr across each layer. The surface energy of each orientation was evaluated (see Supplementary Materials for calculation details). Then, a random distribution of 558 LiF molecules, 504 KF molecules, and 138 NaF molecules (LiF-NaF-KF: 46.2-11.5–42 mol\%), which corresponds to the composition seen in FLiNaK salt studies \cite{williams2006assessment} is generated using the packmol package \cite{martinez2009packmol}. The salt was pre-equilibrated at five temperatures (600, 650, 700, 750, and 800°C) under atmospheric pressure using the Berendsen barostat \cite{berendsen1984molecular} and the Nose-Hoover thermostat \cite{nose1984unified} with a time step of 0.25 fs. The $x$ and $y$ dimensions of the salt were constrained to match the slab dimensions during relaxation. The resulting density is consistent with experimental values \cite{chrenkova2003density,frandsen2020structure}. Each equilibrated salt system was then placed atop the respective NiCr slab for further simulations. The relaxed simulation cells and the dimensions are shown in Figure \ref{fig:initial}.

Following structure construction, production simulations were conducted under the constant volume and temperature condition (NVT) at the five different temperatures using the Nose-Hoover thermostat \cite{nose1984unified}, with periodic boundary conditions and a 0.25 fs time step. The lowermost two layers of the alloy were fixed throughout the simulations to mimic bulk material constraints. To assess the influence of external electric fields, additional simulations were performed at 800°C with a $\pm$0.10 V/Å field applied along the $Z$-direction. A reflecting wall was imposed at the upper boundary of the simulation cell to confine salt molecules. Dynamic atomic charges were calculated using the Electronegativity Equalization Method (EEM) \cite{mortier1985electronegativity}, which allows for real-time charge redistribution at the alloy-salt interface. Note that EEM may face challenges in accurately accounting for polarization effects under external electric fields. However, at low to moderate field strengths, it remains a useful approach for capturing key trends in ionic polarization and transport, as demonstrated in previous studies \cite{koski2021water}. Fluoride coverage on the alloy surface was determined using a cutoff distance of 80\% of the van der Waals radii, yielding Cr-F and Ni-F cutoff distances of 2.776 Å and 2.48 Å, respectively. The number of dissolved atoms was identified as metal atoms with fewer than three neighboring metal atoms, based on their bonding with fluorine and local coordination environment. For each setup, 10 independent simulations were averaged to ensure statistical quality. The OVITO package \cite{stukowski2009visualization} was used for data post-processing.

\section{Results and Discussion}

\subsection{Orientation-dependent Corrosion Kinetics}

\begin{figure}[!ht]
	\centering
	\includegraphics[width=0.9\textwidth]{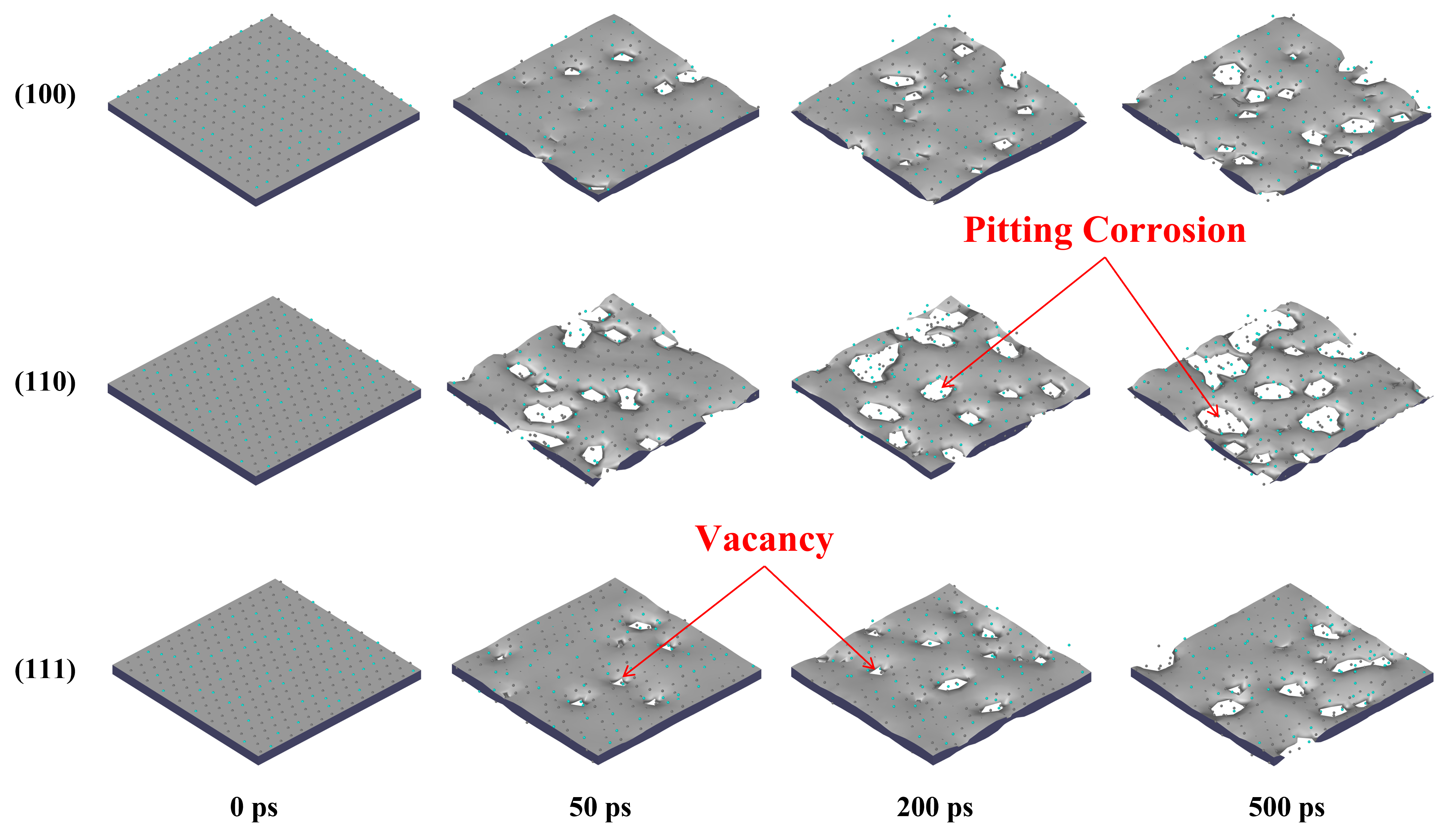}
	\caption{Snapshots of the surface mesh of three NiCr slabs captured at various time frames during the simulation at 700 $\mathrm{^o}$C. The gray regions depict the surface mesh, while the navy blue sections highlight the sides. The arrow indicates holes caused by vacancies from dissolved surface elements, with the larger area representing pitting corrosion.}
	\label{fig:surface}
\end{figure}

The corrosion kinetics of NiCr alloy slabs with (100), (110), and (111) orientations were analyzed at 700°C to assess the influence of crystallographic orientation. Figure \ref{fig:surface} illustrates the surface morphology at 0, 50, 200, and 500 ps. Initially, all surfaces are smooth, but early signs of corrosion appear, particularly on the (110) orientation, which exhibits incipient pitting and surface roughening. By 200 ps, the (110) surface shows significant pitting corrosion, whereas the (100) and (111) surfaces maintain a more intact morphology. At 500 ps, the (110) surface undergoes the most severe degradation, with pitting extending into the deeper atomic layers. In contrast, the (100) and (111) surfaces remain comparatively stable with fewer surface voids, with (111) surface showing the least damage. Additional comparisons for the three surfaces, particularly the changes in the first two atomic layers, are provided in the SM. The higher density of reactive sites and lower atomic packing efficiency of (110) make it more susceptible to dissolution and diffusion-driven restructuring compared to the more stable (100) and (111) surfaces. Initially, as Cr preferentially dissolves, terrace vacancies form, which subsequently migrate and coalesce into lateral vacancy clusters (voids/pits). Although Cr is the dominant dissolving species, the locations of pitting do not necessarily correspond to Cr-rich regions. Instead, rapid surface diffusion promotes vacancy migration, leading to the expansion and deepening of pits beyond their initial formation sites. This behavior suggests that the competition between dissolution kinetics and surface diffusion governs the evolution of corrosion-induced morphology. The continuation of corrosion could lead to the 3D porous structure with Ni-rich surface, as noted in previous dealloying experiments \cite{mccue2016dealloying,liu2021formation,liu2023temperature}. 

\begin{figure}[!ht]
	\centering
	\includegraphics[width=1.0\textwidth]{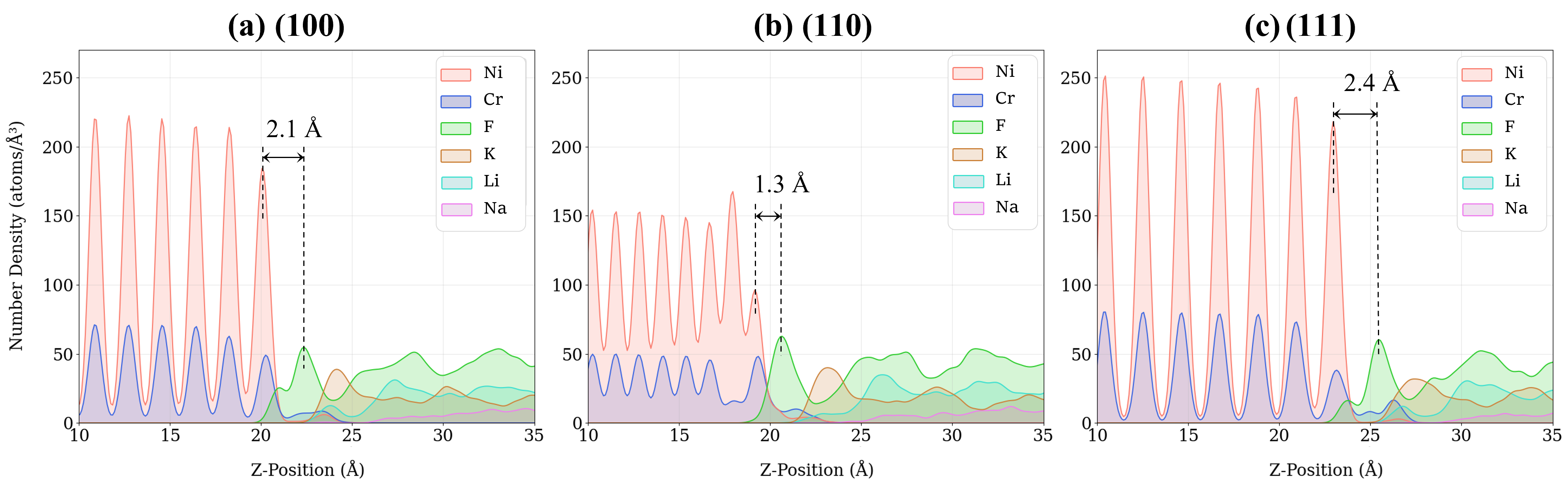}
	\caption{The average atomic density distributions along the $Z$-direction at 500 ps for the simulations of the NiCr (a) (100), (b) (110), and (c) (111) orientations at 700 $\mathrm{^o}$C.}
	\label{fig:density}
\end{figure}

The atomic number density along the slab normal direction is presented in the Figure \ref{fig:density}. The (100) and (111) surfaces retain well-defined Ni and Cr peaks near the surface, indicating limited surface disruption. In contrast, the (110) surface shows the most severe structural disruption, characterized by broadened and diminished density peaks near the surface. A significant depletion of Cr in the second atomic layer is observed, due to outward Cr diffusion to the surface layer. On the other hand, Ni atoms diffuse inward to compensate for Cr loss in the second layer. It leads to the gradual dealloying of the metal surface region with a Ni-rich composition. Such phenomena occur for all the surfaces, though with a significantly higher rate for (110) surface. In contrast, the (100) and (111) surfaces exhibit much less dealloying within the simulation timeframe, with (111) surface showing the least Ni/Cr counter-diffusion. Details of the atomic process are visualized in the SM Figure S2 and Supplementary videos.

\begin{figure}[!ht]
	\centering
	\includegraphics[width=1.0\textwidth]{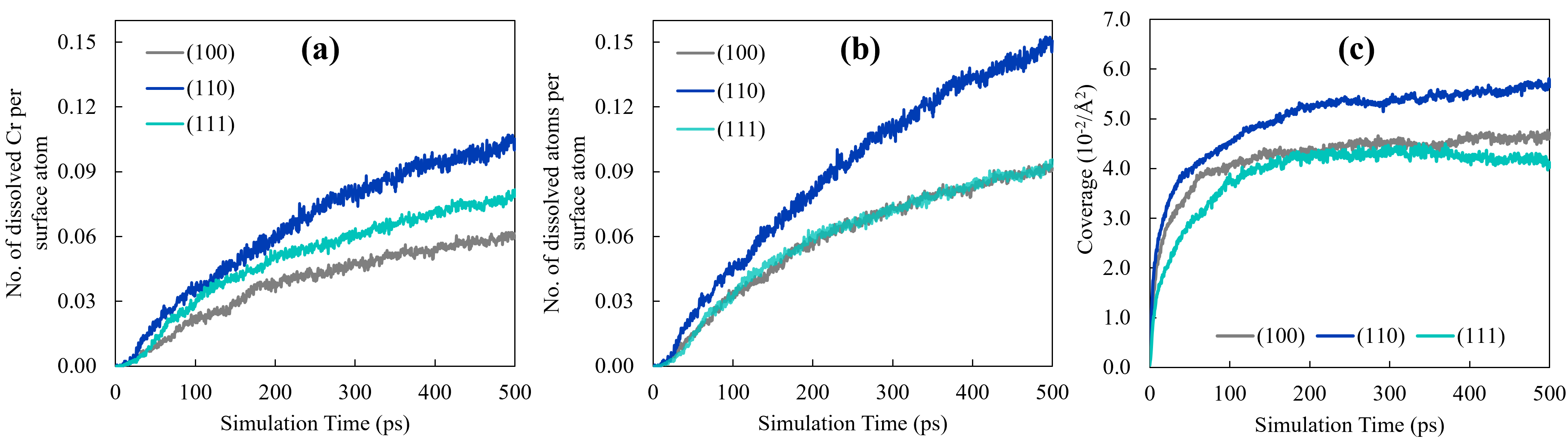}
	\caption{Time evolution of (a) the number of dissolved Cr atoms per surface atom, (b) the total number of dissolved atoms per surface atom, and (c) the fluorine coverage on the metal surfaces.}
	\label{fig:evolution}
\end{figure}

A quantitative assessment of corrosion rate is provided in Figures \ref{fig:evolution}(a-c), which track the evolution of dissolved Cr atoms and total dissolved atoms per surface atom (as the surface atomic density varies for each crystallographic orientation, the number of dissolved atoms is normalized by the number of surface atoms). The (110) surface exhibits the highest Cr dissolution, while the (100) slab shows the lowest. The (111) surface experiences greater Cr loss than (100) but lower Ni dissolution, resulting in nearly identical total dissolved atoms for both orientations. Although the (100) and (111) surfaces show similar overall dissolution levels (Figure \ref{fig:evolution}(b)), Supplementary Figure S1 reveals that the second layer of the (111) surface has fewer vacancies, indicating less subsurface damage than the (100) surface. This suggests that over extended exposure to fluoride salt as expected in realistic conditions, the (111) surface may exhibit the least overall corrosion. The surface energies based on this ReaxFF were computed to be 10.77 eV/nm$^2$, 11.85 eV/nm$^2$, and 10.11 eV/nm$^2$ for the (100), (110), and (111) surfaces, respectively. The higher surface energy of (110) correlates with its greater susceptibility to corrosion, while the lower surface energy of (111) suggests greater morphological stability over time. However, as corrosion progresses, the surface structure deviates from its pristine state, meaning that these energy values provide only a qualitative understanding of stability trends rather than precise predictive metrics. Ultimately, the corrosion behavior is dictated by the relative rates of dynamic processes, including F adsorption, metal dissolution, and surface diffusion.

Previous work by Arkoub et al. \cite{arkoub2024reactive} identified F surface coverage as a key indicator of alloy-salt interaction. Hence, we computed the evolution of F surface coverage on the surface as shown in Figure \ref{fig:evolution}(d); the (110) slab exhibits the highest F coverage, aligning with its greater dissolution rate. In comparison, the (100) and (111) surfaces show lower yet comparable coverage levels, with the (111) surface exhibiting the lower coverage. The double-layer thickness, defined as the distance between the first F peak and the metal peak in the atomic density distribution, is quantified. As shown in Figure \ref{fig:density}, the (110) surface has the smallest double-layer thickness (1.3 Å) compared to 2.1 Å for (100) and 2.4 Å for (111), suggesting stronger F-metal interactions for the (110) surface. Prior DFT-based adsorption energy calculations \cite{yin2018theoretical} yielded F adsorption energy follows the order (110) $>$ (100) $>$ (111) on pure Ni and Cr-doped Ni surfaces. Hence, the adsorption energy is a reasonable indicator of the F distribution over the metal surface under the dynamic conditions.

Note that Ni dissolution occurs in tandem with Cr dissolution but at a much lower magnitude. As the noble metal in this binary alloy, Ni is relatively resistant to fluoride formation \cite{olson2009materials, sridharan2013corrosion}, and its dissolution is primarily triggered by structural destabilization following Cr depletion. Ni dissolution and redeposition were proposed in previous experimental studies \cite{liu2021formation, liu2023temperature}.  A similar behavior was reported by Chen et al. \cite{chen2021effect}, who demonstrated that Ni oxidation does not initiate through oxygen diffusion into the matrix but rather through Ni detachment from the surface to react with adsorbed oxygen, a process facilitated by vacancies. We believe that, in molten salt environments, the vacancies left by dissolved Cr atoms similarly weaken local bonding, enhancing Ni mobility and detachment probability at elevated temperatures. This understanding aligns with our previous work \cite{arkoub2024reactive}, where Ni dissolution only became significant in high Cr content alloys (Ni${_{0.75}}$Cr${_{0.25}}$ in FLiNaK), where Cr dissolution was most intensive. Among the examined crystallographic orientations, the (111) surface exhibits the lowest Ni dissolution over the 500 ps simulation period (Figure \ref{fig:evolution}(b-c)), a trend that correlates with its reduced Ni diffusion, which will be discussed further in subsequent sections. Despite this, Ni dissolution remains low compared to Cr, consistent with experimental results \cite{olson2009materials, mcalpine2020corrosion, arkoub2024reactive}.

\begin{figure}[!ht]
	\centering
	\includegraphics[width=1.0\textwidth]{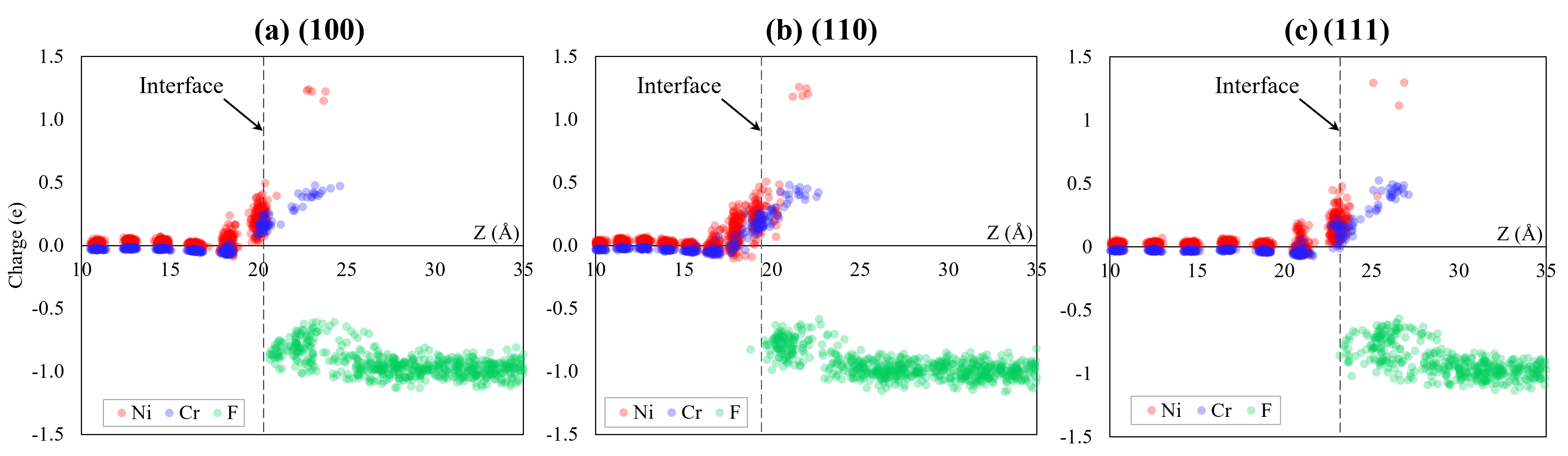}
	\caption{Charge distribution of Ni, Cr, and F atoms along $Z$-direction at 500 ps for the simulation of the NiCr (a) (100), (b) (110), and (c) (111) slab orientations at 700 $\mathrm{^o}$C.}
	\label{fig:charge}
\end{figure}

The charge distribution at the alloy-salt interface reveals significant electron transfer, primarily driven by the strong affinity of F ions to metal atoms. Figure \ref{fig:charge} illustrates the charge profiles of Ni, Cr, and F across the three slab orientations at 500 ps. In the bulk salt, F ions exhibit a charge close to -1, but this charge state diminishes and fluctuates as they approach the interface. These dynamic charge states of interfacial atoms indicate the active electron transfer between the near-surface metal atoms and adjacent F ions, which facilitates metal dissolution. Upon dissolution, both Ni and Cr exhibit increased charge magnitudes, consistent with trends reported by Arkoub et al. \cite{arkoub2024reactive}. By comparison, dissolved Ni retains a significantly higher positive charge than Cr, a difference that is attributed to fluoride complexion and charge delocalization effects. Cr strongly interacts with fluoride ions, forming CrF$_x$ complexes (e.g., CrF$_2$ and CrF$_3$), which delocalize its charge and reduce its net positive state compared to isolated Cr ions. In contrast, Ni forms weaker fluoride complexes \cite{guo2018corrosion}, resulting in a more localized charge state and higher residual positive charge in solution. The (110) slab exhibits the most pronounced charge perturbation, with fluctuations extending into multiple atomic layers, reflecting deeper structural disruption. In contrast, charge variations in the (100) and (111) slabs remain largely confined to the top layer, indicating more localized interactions with the molten salt. These charge distribution trends align closely with the morphological evolution shown in Figure \ref{fig:evolution}(a), where the (110) surface undergoes the most severe degradation, while the (100) and (111) surfaces remain more structurally intact.
 
\subsection{Temperature Effects on Corrosion Kinetics}

\begin{figure}[!ht]
	\centering
	\includegraphics[width=0.8\textwidth]{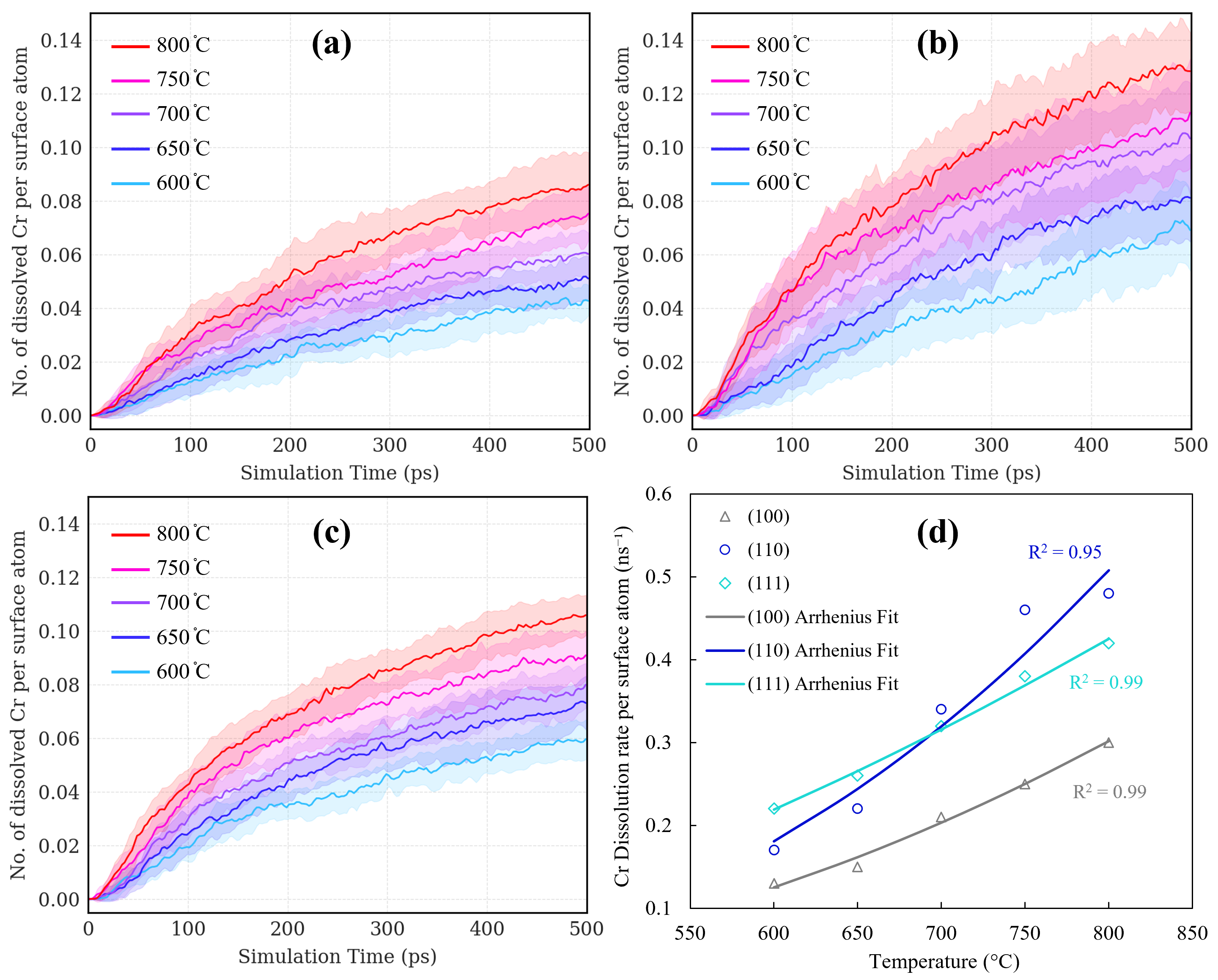}
	\caption{Time evolution of the number of dissolved Cr atoms per surface atom for NiCr alloy slabs at (a) (100), (b) (110), and (c) (111) orientations, simulated over 500 ps across five temperatures: 600 $\mathrm{^o}$C, 650 $\mathrm{^o}$C, 700 $\mathrm{^o}$C, 750 $\mathrm{^o}$C, and 800 $\mathrm{^o}$C. Shaded regions indicate the standard deviation of the dissolved atom counts. (d) Initial Cr dissolution rate per surface atom for the (100), (110), and (111) orientations as a function of temperature, with solid lines representing the Arrhenius equation fits.}
	\label{fig:temperature}
\end{figure}

The effect of temperature on the dissolution and diffusion kinetics of NiCr alloy surfaces with (100), (110), and (111) orientations was investigated across a temperature range of 600 to 800 $\mathrm{^o}$C. As expected, the corrosion rate increases with temperature.  As shown in Figure \ref{fig:temperature}(a-c), the dissolution trend for Cr generally follows the order: (110) $>$ (111) $>$ (100), indicating the (110) surface's greater susceptibility to corrosion across the temperature range. To obtain the temperature-dependent dissolution rates, we focus on the initial linear regime (0–150 ps) of the dissolution curves. This avoids the slowing down of dissolution in the later simulation stage, due to i) as the surface Cr content decreases, the availability of reactive Cr sites decreases, leading to a self-limiting dissolution process where the corrosion rate drops; ii) Cr near the surface reduces the free F due to the formation of CrF$_x$ complexes. The initial dissolution rates are further analyzed using the Arrhenius relationship, which yields the activation energy required for Cr dissolution ($E_d$).  The sampled points and the fitted Arrhenius curves for the three orientations are shown in Figure \ref{fig:temperature}(d). The results indicate that Cr dissolution follows an Arrhenius-type behavior, reflecting the activation-controlled kinetic nature of the corrosion process. Excellent fits were found for the (100) and (111) surfaces ($R^2$ = 0.99), while the (110) slab exhibits slightly more scattering values ($R^2$ = 0.95), which is attributed to faster surface restructuring (Figure \ref{fig:surface}), particularly at higher temperatures.

Based on the Arrhenius fitting, the activation energies for Cr dissolution on the (100), (110), and (111) surfaces are estimated as 0.36, 0.41, and 0.27 eV, respectively. Notably, experimental studies on Hastelloy-B corrosion in FLiNaK at 650, 750, and 850°C \cite{sona2014high} reported an activation energy of 0.36 eV/atom, which aligns closely with the current results. More importantly, these values are significantly lower than the diffusion barrier (0.83 eV) previously reported for Cr diffusion in bulk NiCr alloys \cite{tucker2010ab}. It suggests that corrosion in molten salts is predominantly a kinetically controlled near-surface process, rather than a bulk diffusion-limited phenomenon. In other words, the progression of corrosion does not necessarily have to rely on the long-range bulk diffusion, but can be driven by the near-surface processes.  Liu et al. \cite{liu2021formation} proposed that the continuous removal of Cr is facilitated by the mobility of Ni atoms at the interface. The current results indicate that, as the near-surface structure of the alloy transitions from a crystalline to a more amorphous state, the subsurface Cr atoms gain enhanced mobility, allowing them to diffuse to the surface and dissolve. Simultaneously, Ni atoms diffuse across the surface, smearing morphological disruptions and enabling corrosion to penetrate deeper into the structure without invoking the high-energy bulk diffusion pathway. This interplay between Ni diffusion and Cr depletion may contribute to a self-perpetuating dealloying mechanism, where the surface composition continuously evolves toward a Ni-rich state while Cr dissolution progresses deeper into the material. This mechanism circumvents the large energy barrier associated with long-range Cr diffusion in the crystalline bulk. Furthermore, atomistic simulations of Cr diffusion along grain boundaries (GBs) suggest that Cr mobility along GBs is not significantly enhanced compared to Ni \cite{uberuaga2022effect}. Thus, the progression of intergranular corrosion is likely governed by the same near-surface processes, involving local atomic diffusion and shuffling near the corrosion front, rather than extensive Cr migration along GBs or through the bulk. Another notable observation is the discrepancy between these kinetic barriers and previous static DFT calculations. Specifically, stepwise DFT-based dissolution calculations of NiCr alloy with an intact surface in vacuum or with H$_2$O/F bonding yielded a much higher dissolution barrier ($>$ 1.1 eV) \cite{ke2020dft,arkoub2024first}. This difference highlights the critical role of dynamic interactions in molten salt environments, where alloy-salt interactions and surface disruptions effectively lower the dissolution barriers.

\begin{figure}[!ht]
	\centering
	\includegraphics[width=1.0\textwidth]{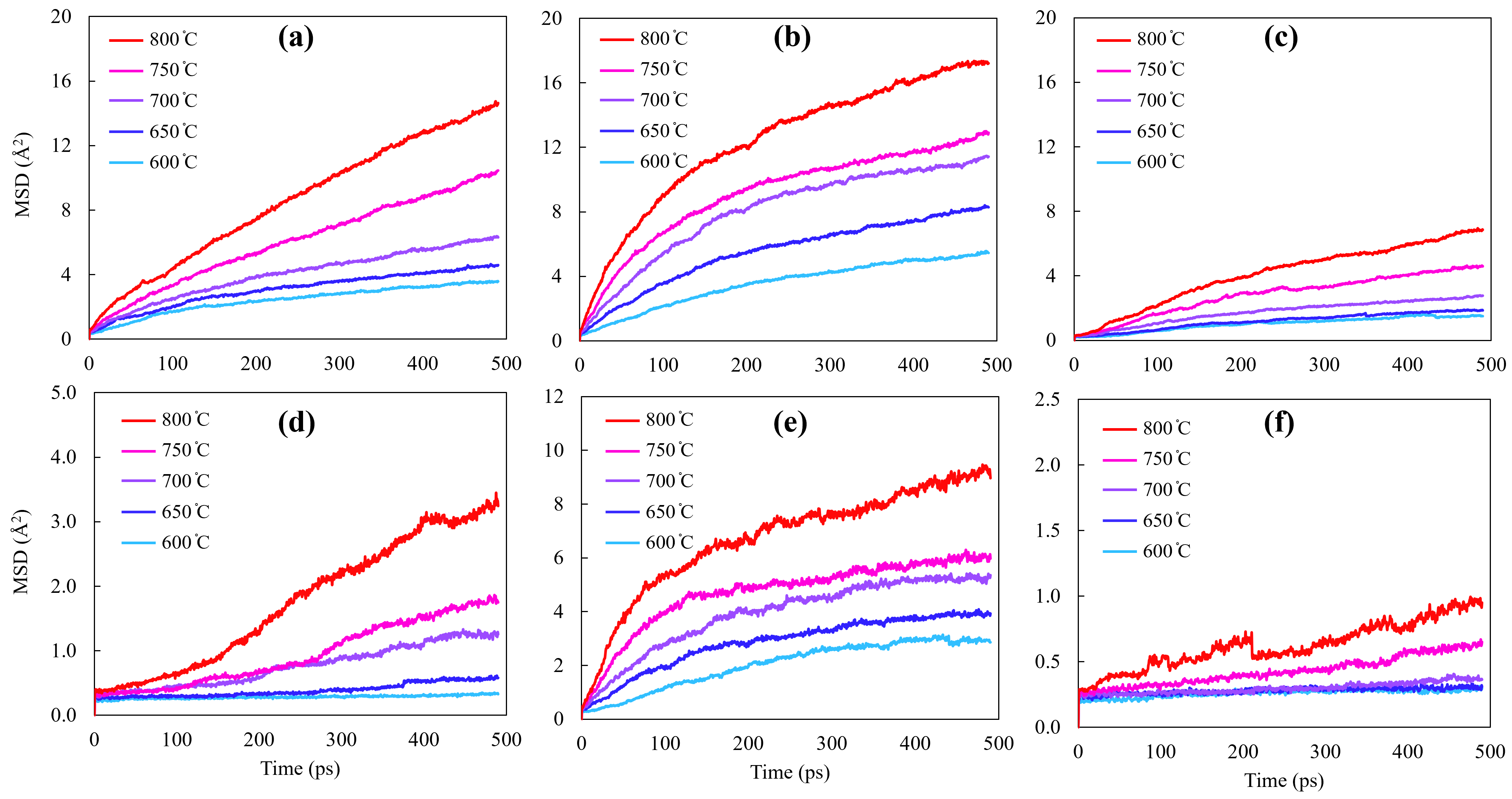}
	\caption{MSD of surface Ni atoms in (a) NiCr(100), (b) (110), and (c) (111) surfaces, as well as Cr atoms in the second layer in (d) NiCr(100), (e) (110), and (f) (111) surfaces, across five temperatures: 600 $\mathrm{^o}$C, 650 $\mathrm{^o}$C, 700 $\mathrm{^o}$C, 750 $\mathrm{^o}$C, and 800 $\mathrm{^o}$C.}
	\label{fig:msd}
\end{figure}

The dissolution trends for Ni, as shown in Figures \ref{fig:evolution}(a-b), present a different order compared to Cr, specifically (110) $>$ (100) $>$ (111). It is speculated that Ni dissolution is likely influenced by the mobility of near-surface atoms, where high diffusivity correlated with greater dissolution susceptibility of Ni. Hence, we computed the mean square displacement (MSD) of top-layer Ni atoms and second-layer Cr atoms, based on the following formulation:
\[\mathrm{MSD}=\frac{1}{N}\langle \sum_{i}^{N} (r_i(t)-r_i(0))^2 \rangle\]
where $N$ is the number of involved atoms, excluding atoms that dissolve into the salt, and $r(t)$ is the atom position at time $t$. As shown in Figure \ref{fig:msd}, the MSD curves reveal a strong dependence of atomic mobility on both temperature and crystallographic orientation. Both (111) and (100) indicate a linear trend in the MSD curves. For both Ni and Cr atoms, the (111) surface consistently exhibits the smallest MSD values. This behavior aligns with the notion that atoms remain tightly bound and resist vacancy formation and migration, which in turn mitigates corrosion progression for the (111) orientation. The (100) surface shows higher MSD magnitudes than the (111) surface; this is attributed to the lower atomic packing density of (100), which increases surface atom mobility. In contrast, the (110) surface exhibits the highest MSD magnitudes with a distinctly nonlinear trend, indicating additional diffusion mechanisms. The initial rapid increase in MSD for both Ni and Cr atoms stems from the simultaneous counter-diffusion, where Cr diffuses outward to the top layer of the alloy-salt interface, while Ni migrates inward to the metal. As the corrosion progresses, the MSD curve transitions to a slower, linear regime, aligning with the (100) and (111) surfaces. This second stage is characterized by dominant lateral diffusion, where atoms move across the surface plane, contributing to surface morphology evolution. Hence, the MSD analysis reinforces the idea that surface diffusion dynamics can be a good indicator of Ni dissolution with (110) $>$ (100) $>$ (111). Overall, the crystallographic orientation plays a crucial role in modulating the surface dynamics, with the (110) surface's high atom mobility contributing to accelerated corrosion, while the (111) surface's stability inhibits material loss. 

\subsection{Electric Field-Assisted Corrosion}

The influence of an external electric field perpendicular to the NiCr surface was examined at 800 $\mathrm{^o}$C. Specifically, two fields with magnitudes $+0.10$ V/Å and $-0.10$ V/Å were applied to explore their effects on the interaction between the NiCr alloy and the FLiNaK salt. The positive field ($+0.10$ V/Å) should, in theory, enhance the attraction of negatively charged species toward the surface, potentially altering the F adsorption behavior and corrosion mechanisms of the system. Conversely, the negative field ($-0.10$ V/Å) will repel anions and may influence the dynamics of cationic species near the surface. By comparing the results, insights can be drawn into how electric fields might be used to control corrosion processes in high-temperature molten salt environments.

\begin{figure}[!ht]
	\centering
	\includegraphics[width=1.0\textwidth]{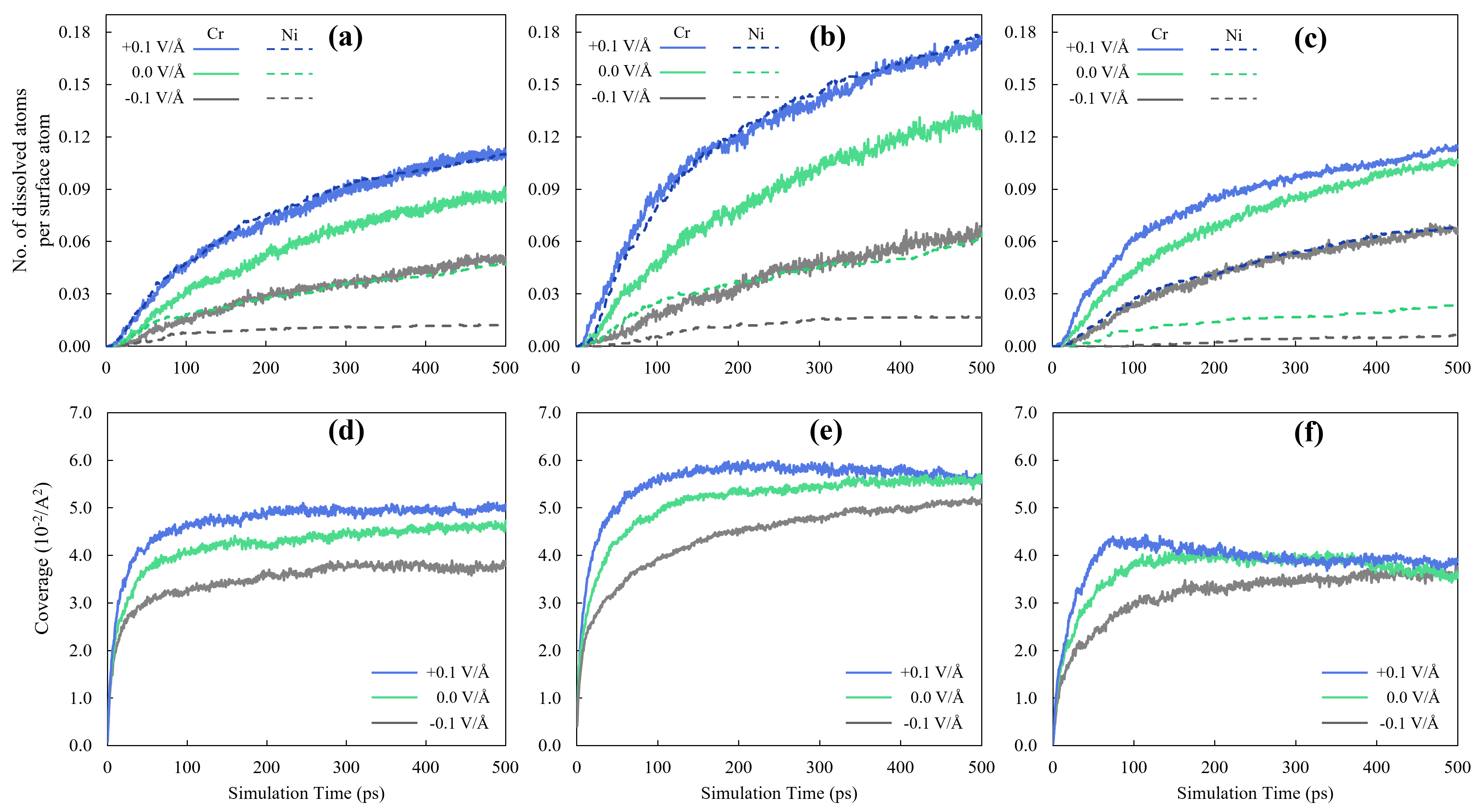}
	\caption{Time evolution of simulations for the (a, d) NiCr(100), (b, e) NiCr(110), and (c, f) NiCr(111) slabs at 800°C under different electric field conditions. Top panels (a–c) show the number of dissolved Cr and Ni atoms in the salt per surface atom, while bottom panels (d–f) depict the fluorine coverage on the surface.}
	\label{fig:evolution_e}
\end{figure}

The time evolution of the number of dissolved atoms and F coverage for the (100), (110), and (111) orientations under external electric fields is presented in Figure \ref{fig:evolution_e}. Compared to the field-free scenario, the application of a positive electric field (+0.10 V/Å) significantly enhances the Cr dissolution rate (Figures \ref{fig:evolution_e}(a-c)). This increase correlates with the enhanced F coverage on the alloy surface (Figures \ref{fig:evolution_e}(d-f)). The positive electric field drives negatively charged F ions toward the NiCr alloy surface, thereby increasing their local concentration at the interface. Such increased concentration has a twofold effect on the alloy surface stability. Firstly, the chemical bonding of F with surface Cr weakens the metallic bonds due to adsorption. Secondly, the physical accumulation of F ions alters the local electric field, inducing surface charge redistribution that destabilizes the surface morphology (to be analyzed later). Consequently, a significantly higher number of Ni atoms are observed to detach from the alloy surface and dissolve into the salt, as depicted in Figures \ref{fig:evolution_e}(a-c).  
 
Conversely, applying a negative electric field of $-0.10$ V/Å exerts a protective effect on the NiCr alloy surface, demonstrating a cathodic corrosion mitigation strategy. As shown in Figures \ref{fig:evolution_e}(a-c), under the negative electric field, the rate of Cr dissolution is substantially reduced, accompanied by a marked decline in F coverage (Figures \ref{fig:evolution_e}(d-f)). The negative electric field establishes an electrostatic repulsion for negatively charged F ions, effectively reducing their local concentration at the interface as indicated in Figures \ref{fig:evolution_e}(d-f). Simultaneously, the negative field pushes positively charged cations (e.g., K$^\mathrm{+}$, Li$^\mathrm{+}$, Na$^\mathrm{+}$) toward the alloy surface, promoting a cationic shielding effect. This redistribution of ionic species induces a local electric double layer, where cation accumulation neutralizes surface charges. It not only reduces the number of F ions adsorbed onto active Cr sites but also modifies the electrochemical potential of the surface, making the alloy less prone to oxidation reactions involving either Cr or Ni.

\begin{figure}[!ht]
	\centering
	\includegraphics[width=1.0\textwidth]{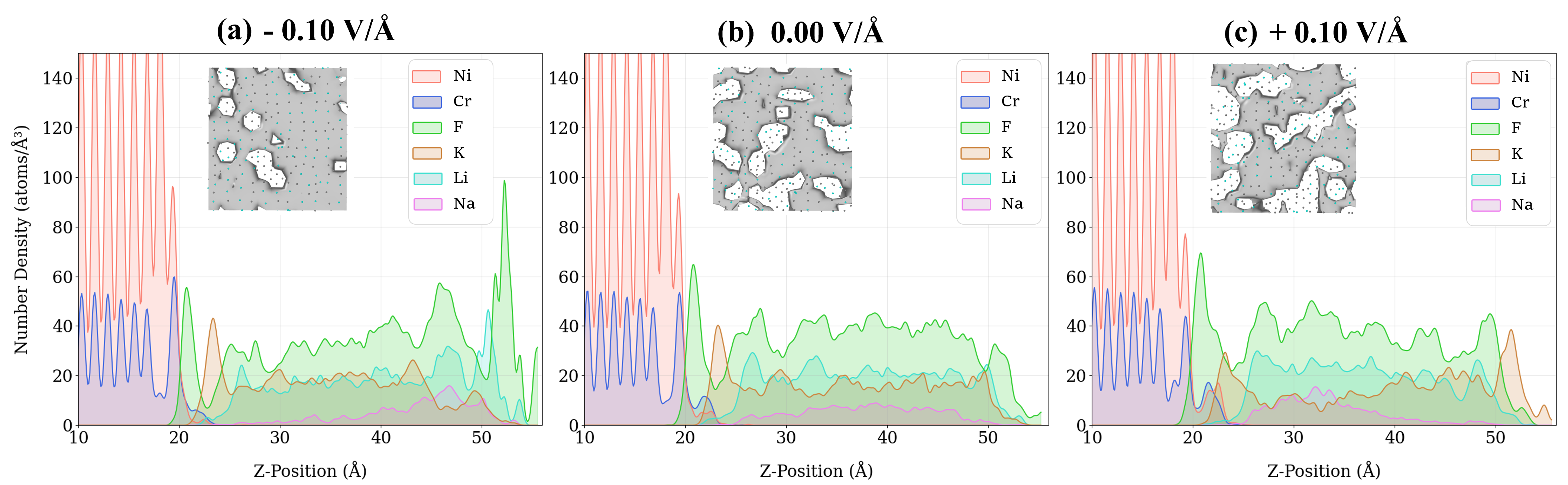}
	\caption{The average atomic density distributions along the $Z$-direction accompanied with surface morphology at 500 ps for the NiCr(110) alloy simulations at 800 $\mathrm{^o}$C under three electric field conditions: (a) -0.1 V/\AA, (b) 0.0 V/\AA, and (c) 0.1 V/\AA. The green dashed arrows highlight the redistribution trend of F ions. The insets show the corresponding surface microstructure.}
	\label{fig:density_e}
\end{figure}

The effect of the electric field on corrosion behavior varies across different crystallographic orientations. When exposed to a positive electric field (+0.10 V/Å), the (110) surface experiences the most pronounced increase in corrosion, whereas the (111) surface exhibits the least increase. Under a negative electric field (-0.10 V/Å), all orientations experience a significant reduction in corrosion. To provide a mechanistic understanding of how electric fields modulate corrosion processes, in the following demonstration, we focus on the NiCr(110) orientation, which has been established as the most corrosion-prone surface in previous discussions and exhibits the most changes under the electrical field. The results for the other two surfaces are included in the SM. The atomic density distributions along the $Z$-direction for the NiCr(110) slab, along with the corresponding surface morphology, at 500 ps are presented in Figure \ref{fig:density_e}. Under the negative electric field (-0.10 V/Å), the F ion density near the surface is substantially reduced, accompanied by an increased presence of cations (e.g., K$^+$ ions) at the interface. The surface indicates a much less disturbed morphology due to the cathodic protection effect. Conversely, the positive electric field (+0.10 V/Å) drives F ions toward the alloy surface, leading to a pronounced F density peak near the surface; this results in accelerated Cr dissolution and altered surface morphology. Interestingly, despite these differences in dissolution rates and surface stability, the thickness of the double-layer at the interface remains almost unchanged under both positive and negative electric fields. This observation implies that the electric field’s primary effect is on ion distribution and local interactions rather than altering the broader electrostatic environment at the surface. 

\begin{figure}[!ht]
	\centering
	\includegraphics[width=1.0\textwidth]{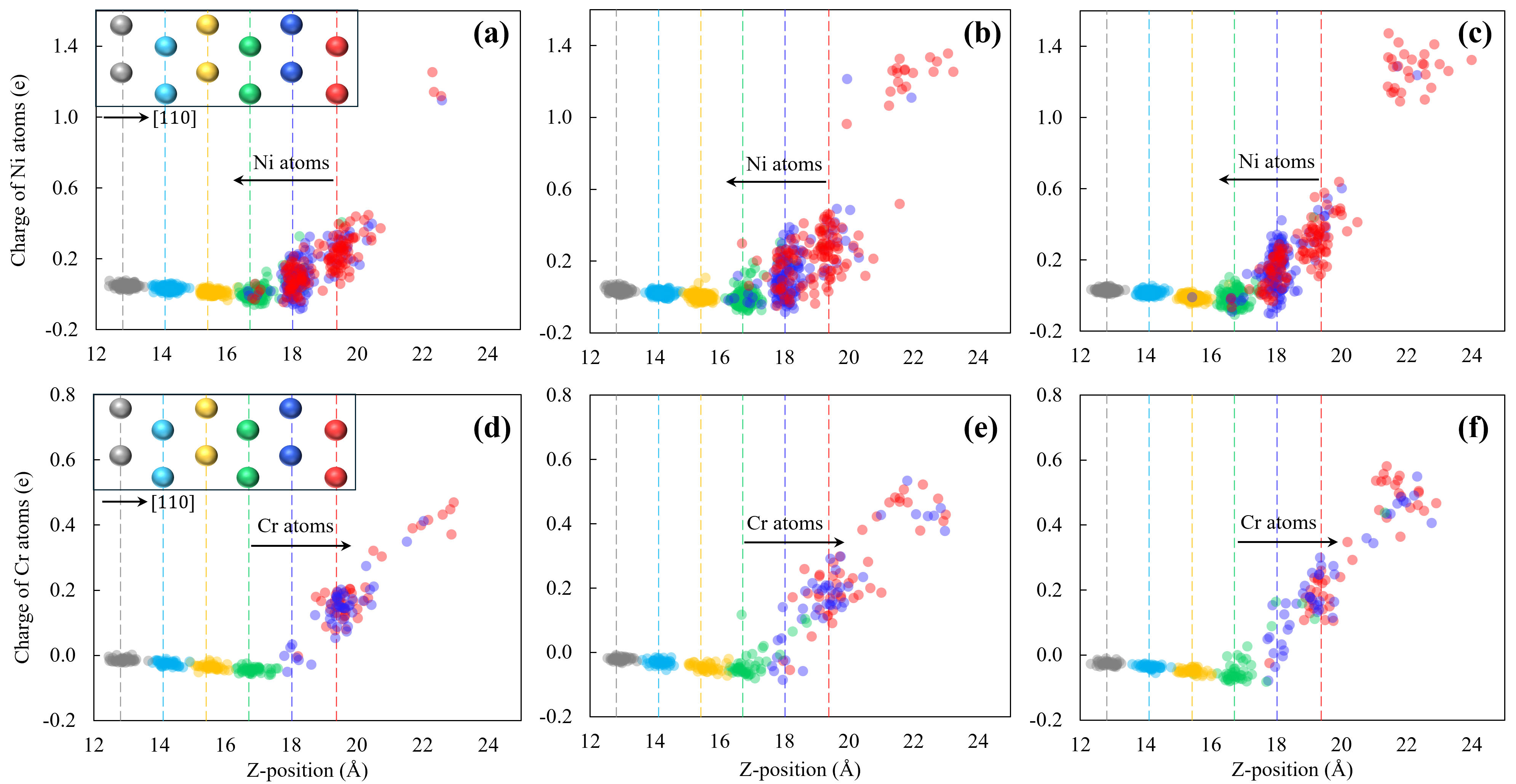}
	\caption{Charge distribution along the $Z$-direction of (a–c) Ni atoms and (d–f) Cr atoms at 500 ps for the NiCr(110) alloy at 800°C under different electric field conditions: (a, d) -0.1 V/Å, (b, e) 0.0 V/Å, and (c, f) 0.1 V/Å. Dashed lines mark the initial atomic layers, and the atom charges are color-coded based on the atoms' layers in the initial perfect structure as shown in the insets of (a) and (d).}
	\label{fig:charge_e}
\end{figure}

The charge distribution of Ni and Cr atoms under different electric field conditions is shown in Figure \ref{fig:charge_e} for the (110) orientation at 500 ps. The atomic charges are color-coded based on the layers the atoms initially belong to, as indicated by the dashed line. The $Z$-coordinate reflects the extent of diffusion and dissolution of Ni and Cr atoms. Regardless of the applied electric field, Ni exhibits a strong inward diffusion tendency, while Cr atoms demonstrate a consistent outward migration. Eventually, it leads to the deeper layers enriched in Ni and depleted in Cr. The influence of a positive electric field (+0.10 V/Å) is particularly evident in Figures \ref{fig:charge_e}(c,f), where Cr atoms from the third layer start to leach. Hence, the positive electric field not only promotes surface Cr dissolution but also accelerates subsurface Cr migration, effectively deepening the corrosion front. In addition, the increased charge disturbance enhances the local charge transfer to drive corrosion reactions, which aligns with the higher dissolution rates observed in Figure \ref{fig:evolution_e}. Regarding the other two orientations, similar field-affected transport behavior was observed (see SM), though the extent of the changes was significantly lower.

\begin{figure}[!ht]
	\centering
	\includegraphics[width=1.0\textwidth]{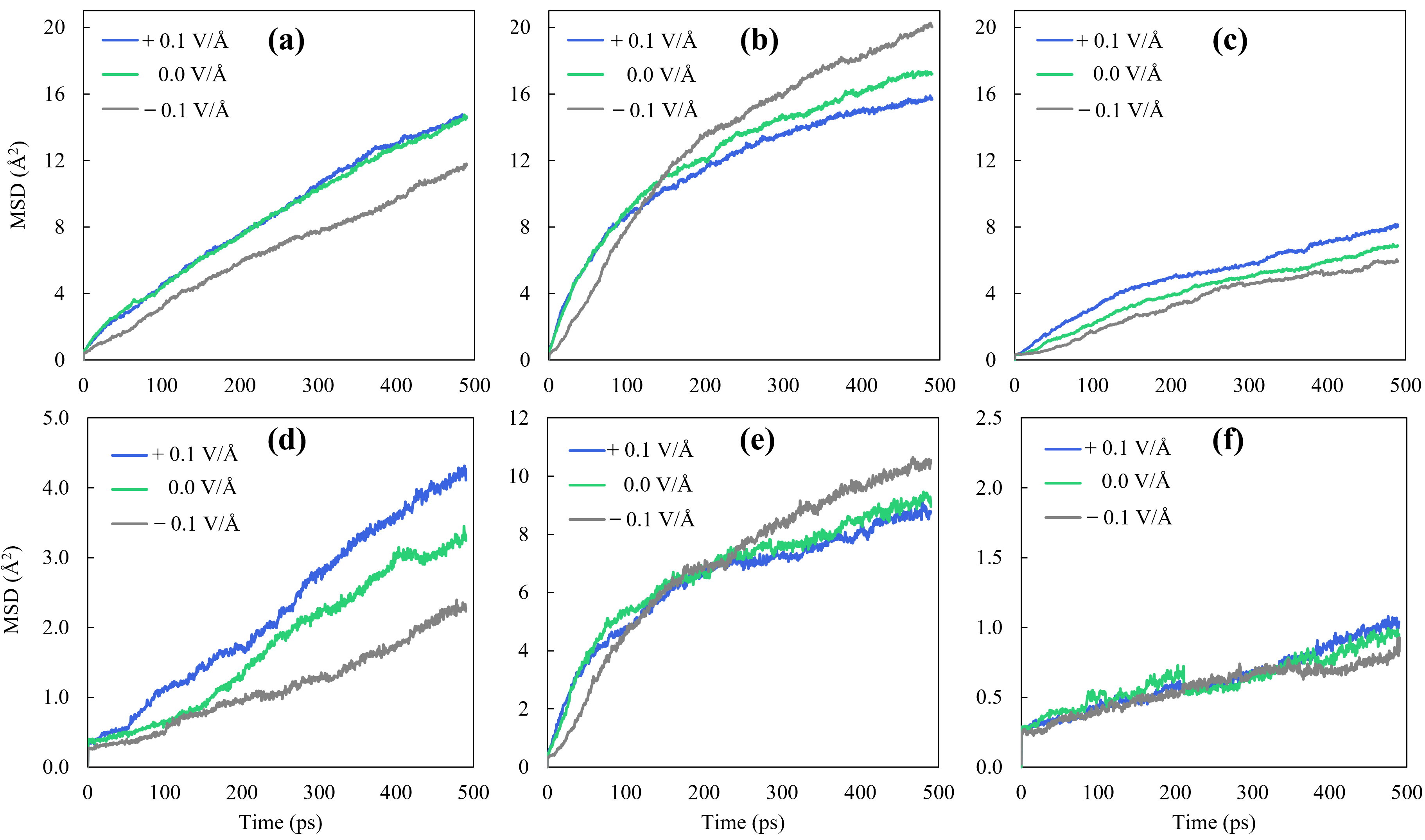}
	\caption{MSD of surface Ni atoms in (a) NiCr(100), (b) (110), and (c) (111) surfaces, as well as Cr atoms in the second layer in (d) NiCr(100), (e) (110), and (f) (111) surfaces, in corrosion kinetics under three electric field conditions: -0.1 V/\AA, 0.0 V/\AA, and 0.1 V/\AA.}
	\label{fig:msd_e}
\end{figure}

The diffusion behavior under external electric fields was analyzed through MSD curves for surface Ni atoms and second-layer Cr atoms across the (100), (110), and (111) orientations, as shown in Figure \ref{fig:msd_e}. The results reveal significant differences in how electric fields influence atomic mobility. For the (111) surface, the MSD values remain consistently low with minimal changes under both positive and negative electric fields. In contrast, the (100) surface shows an asymmetric response to electric fields, where the positive field has little impact on Ni diffusion, but the negative field significantly suppresses atomic mobility for both Ni and Cr atoms. This suppression is attributed to cationic shielding, which assists in stabilizing atomic positions. This mechanism likely accounts for the reduction in MSD for Ni on the (110) surface under a negative electric field in the initial stage. Interestingly, as the system evolves, in the later stage of the (110) surface, the diffusion rate is negatively correlated with Cr dissolution, which is likely due to the combined effects of adsorbed F ions and the concentration of vacant sites. Such enhanced Ni mobility under the negative electric field could contribute to a self-healing effect via accelerated surface passivation for (110) surface. By contrast, the positive electric field does not show any strong effect on Ni and Cr diffusion for the (110) surface. Thus, the electric field, by modifying local interactions, introduces a complex coupling between diffusion and dissolution kinetics. A negative electric field, in particular, can be employed for reducing dissolution rates while enhancing surface self-healing mechanisms, which offers a promising approach for controlling corrosion behavior in high-temperature molten salt environments.   

\section{Conclusion}

This study provides atomistic insights into the orientation-dependent corrosion behavior of NiCr alloys in molten FLiNaK salt using ReaxFF MD simulations. Through the analyses of how crystallographic orientation, temperature, and applied electric fields influence corrosion kinetics, we note the interplay between surface atomic structure and environmental conditions as summarized below,

\begin{itemize}
    \item Among the investigated orientations, the (110) surface exhibits the highest corrosion susceptibility, characterized by rapid Cr dissolution, pronounced surface roughening and pitting. In contrast, the (100) and (111) surfaces demonstrate greater stability, with (111) showing the least material degradation. These differences are linked to the coupling among dissolution, fluoride ion adsorption, and surface diffusion, collectively contributing to the evolution of surface morphology.
    \item The Cr dissolution process follows an Arrhenius-type dependence on temperature, with activation energies deduced to be 0.36, 0.41, and 0.27 eV for the (100), (110), and (111) orientations, respectively. Such experimentally consistent, low dissolution barriers suggest that the progression of the corrosion front is a kinetically controlled near-surface process rather than a long-range bulk diffusion-limited process. The progression of corrosion front is sustained by Ni/Cr counter-diffusion and Cr dissolution from sub-surface amorphous layers.
    \item The application of external electric fields introduces an effective tool for modulating corrosion behavior. A positive electric field (+0.10 V/Å) accelerates Cr dissolution by enhancing fluoride adsorption and destabilizing the surface, while a negative field (-0.10 V/Å) significantly reduces the corrosion. This electrochemical control strategy demonstrates that tailoring electric fields can effectively mitigate corrosion.
\end{itemize}

These insights contribute to a broader understanding of the degradation mechanisms of Ni-based alloys in molten salt environments and suggest pathways for designing more corrosion-resistant materials for molten salt applications.

\section*{Acknowledgments}
This research was funded by the National Science Foundation (NSF) through CAREER Award No. 2340019. The views, findings, conclusions, or recommendations presented in this work are solely those of the authors and do not necessarily represent those of the NSF.

\bibliographystyle{elsarticle-num}
\bibliography{references}

% \begin{thebibliography}{00}

% \bibitem[ ()]{}

% \end{thebibliography}
\end{document}